\documentstyle[aps,eqsecnum,floats]{revtex}

\begin{document}
\draft

\title{Asymptotically constrained and real-valued system based on
Ashtekar's variables\footnote{gr-qc/9906062, to appear in Phys. Rev. D.
(Rapid Communication)}}
\author{Hisa-aki Shinkai  \cite{Email-his}}
\address{
Center for Gravitational Physics and Geometry, 
104 Davey Lab., Department of Physics,
The Pennsylvania State University,
University Park, Pennsylvania 16802-6300, USA
}
\author{Gen Yoneda  \cite{Email-yone}}
\address{
Department of Mathematical Sciences, Waseda University,
Shinjuku, Tokyo,  169-8555, Japan
}
\date{August 16, 1999}
\maketitle
\begin{abstract}
We present a set of dynamical equations
based on Ashtekar's extension of the Einstein equation.
The system
forces the space-time to evolve  to the manifold that satisfies
the constraint equations or the reality conditions or both 
as the attractor 
against perturbative errors.
This is an application of the idea by
Brodbeck, Frittelli, H\"ubner and Reula who constructed
an asymptotically stable (i.e., constrained) system  for the Einstein 
equation, adding 
dissipative forces in the extended space.
The obtained systems may be useful for future numerical studies using
Ashtekar's variables.

\end{abstract}



\section{Introduction} \label{sec:intro}

Ashtekar's extended formulation of general relativity
\cite{Ashtekar} has many
advantages.
By using his special pair of variables,
the constraint equations which appear in the formulation become
low-order polynomials, and the theory has the
correct form for gauge theoretical interpretation. These features
suggest the possibility for developing a nonperturbative quantum
description of gravity.
When we apply his formulation to classical dynamics,
however,
we have to impose the reality condition additionally to the system,
since Ashtekar's variables will not produce a 
real-valued metric in general.

Fortunately, it was shown that the secondary condition of the reality
of the metric will
be automatically preserved during the evolution,
if the initial data satisfies
both primary and secondary metric reality conditions
\cite{AshtekarRomanoTate}.
If we impose the reality condition on the triad 
(triad reality condition), then we have additional conditions
that can be controlled by a part of a gauge variable, triad lapse
${\cal A}^a_0$ (defined later) \cite{ys-con}.
Therefore the reality conditions are controllable,
and we think that
applying the Ashtekar formulation to dynamics is
quite attractive, and  broadens our
possibilities to attack dynamical issues.

It is, however, also the case that preliminary
 numerical simulations of the
spacetime using Ashtekar's variables show that
the system will not normally recover real-valued spacetime
if we relax the metric reality condition locally during the
evolution \cite{ysn-dege}.
Therefore, we desire a system that  is robust for controlling
both the
reality conditions and the constraint equations for stable
long-term integration. 

In this article, we propose new dynamical systems
based on Ashtekar's
variables, which satisfy this desire.
That is, even if numerical data gives us a truncated solution
which violates the constraints or the reality conditions
during steps of time evolution,
the system
forces the spacetime to evolve to a manifold that satisfies
the constraint equations and the reality conditions against these
small deviations.
This is an extension of the idea by
Brodbeck, Frittelli, H\"ubner and Reula (BFHR) \cite{BFHR}
who constructed
an asymptotically {\it stable}  system
(i.e.,  it approaches to the constraint surface) for
the Einstein equation. (A similar effort can be found also
in \cite{detweiler}.)
BFHR introduced  additional dynamical variables, $\lambda$s,
which obey dissipative
dynamical equations and which evolve the spacetime
to the constraint surface of general
relativity as the attractor in the extended spacetime.

Recently, we have obtained a set of equations of motions for
Ashtekar's canonical variables ($\tilde{E}^i_a, {\cal A}^a_i$) in a
symmetric hyperbolic form \cite{YShypPRL} 
(see also \cite{Iriondo,ILRsecond}).
We here present first that a set of constraint equations
also forms a symmetric hyperbolic system in its
evolution equation. 
This is the Ashtekar version of the work by Frittelli
\cite{Fri-con}, and is already shown by Iriondo {\it et al.}
\cite{ILRsecond}.  Our version, however, is based on the fixed
inner product throughout evolution.
We then show that a dissipative system for the combination of
($\tilde{E}^i_a, {\cal A}^a_i$) and constraints also forms a symmetric
hyperbolic system, following the procedure of {\it $\lambda$-system}
by BFHR.
We expect that this system has an attractor in the constraint
surfaces.

We next extend this system also to  one that has an attractor
in the real-valued surfaces.
Since the dynamical system that we have obtained
in a symmetric hyperbolic form
requires the triad reality condition \cite{YShypPRL}, our
purpose is to construct a system which asymptotically evolves into
the triad-real-valued manifold.
We show this is available by applying the same technique.


In this article, we discuss only the case of a vacuum spacetime,
but including matter is straightforward.

\section{Ashtekar formulation}\label{sec:ash}
We start by giving a brief review of the Ashtekar formulation
and the way of handling reality conditions.
We also describe a symmetric hyperbolic system that was
obtained in \cite{YShypPRL}. 

\subsection{Variables and Equations}
The key feature of  Ashtekar's formulation of general relativity
\cite{Ashtekar} is the introduction of a self-dual
connection as one of the basic dynamical variables.
Let us write
the metric $g_{\mu\nu}$ using the tetrad
$E^I_\mu$ as $g_{\mu\nu}=E^I_\mu E^J_\nu \eta_{IJ}$
\footnote{We use
$\mu,\nu=0,\cdots,3$ and
$i,j=1,\cdots,3$ as spacetime indices, while
$I,J=(0),\cdots,(3)$ and
$a,b=(1),\cdots,(3)$ are $SO(1,3)$, $SO(3)$ indices respectively.
We raise and lower
$\mu,\nu,\cdots$ by $g^{\mu\nu}$ and $g_{\mu\nu}$
(the Lorentzian metric);
$I,J,\cdots$ by $\eta^{IJ}={\rm diag}(-1,1,1,1)$ and $\eta_{IJ}$;
$i,j,\cdots$ by $\gamma^{ij}$ and $\gamma_{ij}$ (the 3-metric);
$a,b,\cdots$ by $\delta^{ab}$ and $\delta_{ab}$.
We also use volume forms $\epsilon_{abc}$:
$\epsilon_{abc} \epsilon^{abc}=3!$.}.
Define its inverse, $E^\mu_I$, by
$E^\mu_I:=E^J_\nu g^{\mu\nu}\eta_{IJ}$ and we impose 
$E^0_a=0$ as the gauge condition. 
We define SO(3,C) self-dual and anti self-dual
connections
$
{}^{\pm\!}{\cal A}^a_{\mu}
:= \omega^{0a}_\mu \mp ({i / 2}) \epsilon^a{}_{bc} \, \omega^{bc}_\mu,
$
where $\omega^{IJ}_{\mu}$ is a spin connection 1-form (Ricci
connection), $\omega^{IJ}_{\mu}:=E^{I\nu} \nabla_\mu E^J_\nu.$
Ashtekar's plan is to use  only the self-dual part of
the connection
$^{+\!}{\cal A}^a_\mu$
and to use its spatial part $^{+\!}{\cal A}^a_i$
as a dynamical variable.
Hereafter,
we simply denote $^{+\!}{\cal A}^a_\mu$ as ${\cal A}^a_\mu$.

The lapse function, $N$, and shift vector, $N^i$, both of which we
treat as real-valued functions,
are expressed as $E^\mu_0=(1/N, -N^i/N$).
This allows us to think of
$E^\mu_0$ as a normal vector field to $\Sigma$
spanned by the condition $t=x^0=$const.,
which plays the same role as that of
Arnowitt-Deser-Misner (ADM) formulation.
Ashtekar  treated the set  ($\tilde{E}^i_{a}$, ${\cal A}^a_{i}$)
as basic dynamical variables, where
$\tilde{E}^i_{a}$ is an inverse of the densitized triad
defined by
$
\tilde{E}^i_{a}:=e E^i_{a},
$
where $e:=\det E^a_i$ is a density.
This pair forms the canonical set.

In the case of pure gravitational spacetime,
the Hilbert action takes the form
\begin{eqnarray}
S&=&\int {\rm d}^4 x
[ (\partial_t{\cal A}^a_{i}) \tilde{E}^i_{a}
+(i/2) \null \! \mathop {\vphantom {N}\smash N}
\limits ^{}_{^\sim}\!\null \tilde{E}^i_a
\tilde{E}^j_b F_{ij}^{c} \epsilon^{ab}{}_{c}
\nonumber \\&&
-
e^2
\Lambda \null \! \mathop {\vphantom {N}\smash N}
\limits ^{}_{^\sim}\!\null
-N^i F^a_{ij} \tilde{E}^j_a
+{\cal A}^a_{0} \, {\cal D}_i \tilde{E}^i_{a} ],
 \label{action}
\end{eqnarray}
where
$\null \! \mathop {\vphantom {N}\smash N}
\limits ^{}_{^\sim}\!\null := e^{-1}N$,
${F}^a_{\mu\nu}
:=
2 \partial_{[\mu} {\cal A}^a_{\nu]}
 - i \epsilon^{a}{}_{bc} \, {\cal A}^b_\mu{\cal A}^c_\nu
$
is the curvature 2-form,
$\Lambda$
is the cosmological constant,
${\cal D}_i \tilde{E}^j_{a}
    :=\partial_i \tilde{E}^j_{a}
-i \epsilon_{ab}{}^c  \, {\cal A}^b_{i}\tilde{E}^j_{c}$,
 and
$e^2=\det\tilde{E}^i_a
=(\det E^a_i)^2$ is defined to be
$\det\tilde{E}^i_a=
(1/6)\epsilon^{abc}
\null\!\mathop{\vphantom {\epsilon}\smash \epsilon}
\limits ^{}_{^\sim}\!\null_{ijk}\tilde{E}^i_a \tilde{E}^j_b
\tilde{E}^k_c$, where
$\epsilon_{ijk}:=\epsilon_{abc}E^a_i E^b_j E^c_k$
 and $\null\!\mathop{\vphantom {\epsilon}\smash \epsilon}
\limits ^{}_{^\sim}\!\null_{ijk}:=e^{-1}\epsilon_{ijk}$
\footnote{When $(i,j,k)=(1,2,3)$,
we have
$\epsilon_{ijk}=e$,
$\null\!\mathop{\vphantom {\epsilon}\smash \epsilon}
\limits ^{}_{^\sim}\!\null_{ijk}=1$,
$\epsilon^{ijk}=e^{-1}$, and
$\tilde{\epsilon}^{ijk}=1$.}.

Varying the action with respect to the non-dynamical variables
$\null \!
\mathop {\vphantom {N}\smash N}\limits ^{}_{^\sim}\!\null$,
$N^i$
and ${\cal A}^a_{0}$ yields the constraint equations,
\begin{eqnarray}
{\cal C}_{H} &:=&
 (i/2)\epsilon^{ab}{}_c \,
\tilde{E}^i_{a} \tilde{E}^j_{b} F_{ij}^{c}
  -\Lambda \, \det\tilde{E}
   \approx 0, \label{c-ham} \\
{\cal C}_{M i} &:=&
  -F^a_{ij} \tilde{E}^j_{a} \approx 0, \label{c-mom}\\
{\cal C}_{Ga} &:=&  {\cal D}_i \tilde{E}^i_{a}
 \approx 0.  \label{c-g}
\end{eqnarray}
The equations of motion for the dynamical variables
($\tilde{E}^i_a$ and ${\cal A}^a_i$) are
\begin{eqnarray}
\partial_t {\tilde{E}^i_a}
&=&-i{\cal D}_j( \epsilon^{cb}{}_a  \, \null \!
\mathop {\vphantom {N}\smash N}\limits ^{}_{^\sim}\!\null
\tilde{E}^j_{c}
\tilde{E}^i_{b})
+2{\cal D}_j(N^{[j}\tilde{E}^{i]}_{a})
\nonumber \\&&
+i{\cal A}^b_{0} \epsilon_{ab}{}^c  \, \tilde{E}^i_c,  \label{eqE}
\\
\partial_t {\cal A}^a_{i} &=&
-i \epsilon^{ab}{}_c  \,
\null \! \mathop {\vphantom {N}\smash N}
\limits ^{}_{^\sim}\!\null \tilde{E}^j_{b} F_{ij}^{c}
+N^j F^a_{ji} +{\cal D}_i{\cal A}^a_{0}+\Lambda \null \!
\mathop {\vphantom {N}\smash N}\limits ^{}_{^\sim}\!\null
\tilde{E}^a_i,
\label{eqA}
\end{eqnarray}
\noindent
where
${\cal D}_jX^{ji}_a:=\partial_jX^{ji}_a-i
 \epsilon_{ab}{}^c {\cal A}^b_{j}X^{ji}_c,$
 for $X^{ij}_a+X^{ji}_a=0$.

\subsection{Reality conditions}
In order to construct the metric  from the variables
$(\tilde{E}^i_a, {\cal A}^a_i,  \null \!
\mathop {\vphantom {N}\smash N}\limits ^{}_{^\sim}\!\null, N^i)$,
we first prepare the 
tetrad $E^\mu_I$ as
$E^\mu_{0}=({1 / e \null \! \mathop {\vphantom {N}\smash N}
\limits ^{}_{^\sim}\!\null}, -{N^i / e \null \!
\mathop {\vphantom {N}\smash N}\limits ^{}_{^\sim}\!\null})$ and
$E^\mu_{a}=(0, \tilde{E}^i_{a} /e).$
Using them, we obtain the  metric $g^{\mu\nu}$ such that
$
g^{\mu\nu}:=E^\mu_{I} E^\nu_{J} \eta^{IJ}. 
$

This metric, 
in general, is not real-valued
in the Ashtekar
formulation.
To ensure that the metric is real-valued,
we need to impose real lapse and shift vectors together with
two {\it metric reality} conditions;
\begin{eqnarray}
{\rm Im} (\tilde{E}^i_a \tilde{E}^{ja} ) &=& 0, \label{w-reality1} \\
W^{ij}:= {\rm Re} (\epsilon^{abc}
\tilde{E}^k_a \tilde{E}^{(i}_b {\cal D}_k \tilde{E}^{j)}_c)
&=& 0,
\label{w-reality2-final}
\end{eqnarray}
where the latter comes from the secondary condition of reality
of the metric
${\rm Im} \{ \partial_t(\tilde{E}^i_a \tilde{E}^{ja} ) \} = 0$
\cite{AshtekarRomanoTate}, and
we assume
$\det\tilde{E}>0$ (see \cite{ys-con}).

For later convenience, we also prepare
stronger reality conditions, {\it triad reality} conditions.
The primary and secondary conditions are written respectively as
\begin{eqnarray}
U^i_a := {\rm Im} (\tilde{E}^i_a ) &=& 0, 
\label{s-reality1} \\
{\rm and~~}
{\rm Im}  ( \partial_t {\tilde{E}^i_a} ) &=& 0.
\label{s-reality2}
\end{eqnarray}
\noindent
Using the equations of motion of $\tilde{E}^i_{a}$,
the gauge constraint (\ref{c-g}),
the metric reality conditions
(\ref{w-reality1}), (\ref{w-reality2-final})
and the primary condition (\ref{s-reality1}),
we see  that  (\ref{s-reality2}) is equivalent to \cite{ys-con}
\begin{eqnarray}
{\rm Re}({\cal A}^a_{0}) &=&
\partial_i( \null \! \mathop {\vphantom {N}\smash N}
\limits ^{}_{^\sim}\!\null )\tilde{E}^{ia}
+(1 /2e) E^b_i \null \! \mathop {\vphantom {N}\smash N}
\limits ^{}_{^\sim}\!\null
\tilde{E}^{ja} \partial_j\tilde{E}^i_b
\nonumber \\&&
+N^{i} {\rm Re}({\cal A}^a_i), \label{s-reality2-final}
\end{eqnarray}
or with un-densitized variables,
\begin{equation}
{\rm Re}({\cal A}^a_{0})=
\partial_i( N)
{E}^{ia}
+N^{i} \, {\rm Re}({\cal A}^a_i).
\label{s-reality2-final2}
\end{equation}
{}From this expression we see that
the secondary triad reality condition
restricts the three components of the ``triad lapse" vector
${\cal A}^a_{0}$.
Therefore (\ref{s-reality2-final}) is
not a restriction on the dynamical variables
($\tilde{E}^i_a $ and ${\cal A}^a_i$)
but on the slicing, which we should impose on each hypersurface.

Throughout the discussion in this article, 
we assume that the initial data of
$(\tilde{E}^i_a, {\cal A}^a_i)$ for evolution are solved so as
to satisfy all three constraint equations and the metric
reality condition (\ref{w-reality1}) and (\ref{w-reality2-final}).
Practically, this is
obtained, for
example, by solving ADM constraints and by transforming a
set of initial data to Ashtekar's notation.

\subsection{A symmetric hyperbolic form}

We say that the system is a first-order (quasi-linear)
partial differential equation system,
if
a certain set of
(complex) variables $u_\alpha$ $(\alpha=1,\cdots, n)$
forms
\begin{equation}
\partial_t u_\alpha
= {\cal M}^{l}{}^{\beta}{}_\alpha (u) \, \partial_l u_\beta
+{\cal N}_\alpha(u),
\label{def}
\end{equation}
where ${\cal M}$ (the characteristic matrix) and
${\cal N}$ are functions of $u$
but do not include any derivatives of $u$.
If the characteristic matrix is a Hermitian matrix, then we say
(\ref{def}) is a symmetric hyperbolic system.

For a pair of $u^{(D)}_\alpha=(\tilde{E}^i_a, {\cal A}^a_i)$,
a symmetric hyperbolic system is obtained by modifying the
right-hand-side of the dynamical equations using appropriate
combinations of the constraint equations.
The final form of the system \cite{YShypPRL} is written as
\begin{eqnarray}
{\cal M}(\tilde{E},\tilde{E})^{labij}&=&
i\epsilon^{abc}  
\null \! \mathop {\vphantom {N}\smash N} \limits ^{}_{^\sim}\!\null
\tilde{E}^l_c \gamma^{ij}
+N^l\gamma^{ij} \delta^{ab},
\label{fm-A}
\\
{\cal M}(\tilde{E},{\cal A})^{labij}&=&{\cal M}({\cal A},\tilde{E})^{labij}=0,
\label{fm-BC}
\\
{\cal M}({\cal A},{\cal A})^{labij}&=&i
\null \! \mathop {\vphantom {N}\smash N} \limits ^{}_{^\sim}\!\null
(\epsilon^{abc} \tilde{E}^j_c \gamma^{li}
- \epsilon^{abc} \tilde{E}^l_c \gamma^{ji}
\nonumber \\ &~&
-e^{-2} \tilde{E}^{ia} \epsilon^{bcd} \tilde{E}^j_c \tilde{E}^l_d
-e^{-2}\epsilon^{acd} \tilde{E}^i_d \tilde{E}^l_c  \tilde{E}^{jb}
\nonumber \\ &~&
+e^{-2}
\epsilon^{acd} \tilde{E}^i_d \tilde{E}^j_c \tilde{E}^{lb}
)
+N^l \delta^{ab} \gamma^{ij},
\label{fm-D}
\end{eqnarray}
where ${\cal M}(\ast,\ast)$ means a block component
of the
characteristic matrix as
\begin{equation}
\partial_t \left[ \begin{array}{l}
\tilde{E}^i_a \\
{\cal A}^a_i
\end{array} \right] \cong
\left[ \begin{array}{cc}
{\cal M}(\tilde{E},\tilde{E})^l {}_a {}^{bi}{}_j &
{\cal M}(\tilde{E},{\cal A})^l{}_{ab}{}^{ij} \\
{\cal M}({\cal A},\tilde{E})^{lab}{}_{ij} &
{\cal M}({\cal A},{\cal A})^{la}{}_{bi}{}^j
\end{array} \right]
\partial_l
\left[ \begin{array}{l}
\tilde{E}^j_b \\
{\cal A}^b_j
\end{array} \right],
\label{matrixform}
\end{equation}
where
$\cong$ means that we have extracted only the terms which
appear in the principal part of the system. The inner product of
a set of the variables is
\begin{equation}
\langle
(\tilde{E}^i_a,{\cal A}^a_i)|
(\tilde{E}^i_a,{\cal A}^a_i)
\rangle
=
\gamma_{ij}\delta^{ab}\tilde{E}^i_a \bar{\tilde{E}}{}^j_b
+
\gamma^{ij}\delta_{ab}{\cal A}^a_i \bar{{\cal A}}{}^b_j.
\end{equation}
We note that this symmetric hyperbolic system is obtained under the
assumption of the triad reality condition, together with gauge
conditions, ${\cal A}^a_0={\cal A}^a_iN^i$ and  $\partial_i N=0$.

\section{Asymptotically constrained system}\label{sec:3}
Frittelli \cite{Fri-con} showed that the propagation of the
constraint equations in the standard ADM system of
the Einstein equation forms a symmetric hyperbolic system.
This fact suggests that a small violation of the constraint
equations such as a truncation error in numerical simulation
does not behave in a fatal way immediately.

Similarly, we can show the set of constraint equations (\ref{c-ham}),
(\ref{c-mom}) and (\ref{c-g}),
forms a symmetric hyperbolic system in its
evolution equations.
The principal part of the time derivatives of ${\cal C}_H$,
$\tilde{{\cal C}}_{Mi} := e \, {\cal C}_{Mi} $ and ${\cal C}_{Ga}$ become
\begin{equation}
\partial_t
\left(
\matrix{
{\cal C}_H \cr \tilde{{\cal C}}_{Mi} \cr{\cal C}_{Ga}}
\right)
\cong
\left(\matrix
{N^l & -e 
\null \! \mathop {\vphantom {N}\smash N} \limits ^{}_{^\sim}\!\null
\gamma^{li} & 0 \cr
-e 
\null \! \mathop {\vphantom {N}\smash N} \limits ^{}_{^\sim}\!\null
\delta_i{}^l &
N^l\delta_i{}^j+i
\null \! \mathop {\vphantom {N}\smash N} \limits ^{}_{^\sim}\!\null
\tilde{\epsilon}{}^{lj}{}_i & 0 \cr
0 & 0 & i
\null \! \mathop {\vphantom {N}\smash N} \limits ^{}_{^\sim}\!\null
\epsilon_a{}^{bc} \tilde{E}^l_c +N^l\delta_a{}^b}
\right)
\partial_l
\left(
\matrix{
{\cal C}_H \cr \tilde{{\cal C}}_{Mj} \cr{\cal C}_{Gb}}
\right),
\label{ct}
\end{equation}
which forms a Hermitian matrix under the inner
product rule of
\begin{eqnarray}
&&
\langle
({\cal C}_H,{\cal C}_{Mi},{\cal C}_{Ga})
|
({\cal C}_H,{\cal C}_{Mi},{\cal C}_{Ga})
\rangle
:=
\nonumber \\ &&
{\cal C}_H \bar{{\cal C}}_H
+\gamma^{ij}{\cal C}_{Mi}\bar{{\cal C}}_{Mj}
+\delta^{ab}{\cal C}_{Ga}\bar{{\cal C}}_{Gb}.
\label{inner}
\end{eqnarray}
We note that non-principal parts of the dynamical equations of a set of
${u}^{(C)}_\alpha=({\cal C}_H, \tilde{{\cal C}}_{M}, {\cal C}_{G})$
include the terms of ${u}^{(D)}_\alpha$ and ${u}^{(C)}_\alpha$. 
These facts suggest that all constraint equations have
a well-posed feature.
Iriondo {\it et al} \cite{ILRsecond} present a similar
result, but our definition of the inner product does not
include any coefficients. We also remark that 
all other occurences
of the inner products throughout this article 
also do not include any coefficients (i.e.,
obey the normal index notation).
Thus we omit to express
the inner product hereafter. 

Following the BFHR procedure \cite{BFHR},
we next construct a dynamical system which evolves the
spacetime to the constrained surface, ${\cal C}_H \approx {\cal C}_{Mi}
\approx {\cal C}_{Ga} \approx 0$ as the attractor.
We introduce new variables ($\lambda, \lambda_i, \lambda_a$),
as they obey the dissipative evolution equations
\begin{eqnarray}
\partial_t\lambda &=&
\alpha_1 \,{\cal C}_H
-\beta_1 \, \lambda, \label{lambdaC1}
\\
\partial_t\lambda_i &=&
 \alpha_2 \,\tilde{{\cal C}}_{Mi}
 -\beta_2 \,\lambda_i, \label{lambdaC2}
\\
\partial_t\lambda_a &=&
\alpha_3 \, {\cal C}_{Ga}
-\beta_3 \, \lambda_a, \label{lambdaC3}
\end{eqnarray}
where $\alpha_i \neq 0$ (allowed to be complex numbers) and $\beta_i > 0$
(real numbers) are constants.

If we take
${u}^{(DL)}_\alpha=(\tilde{E}^i_a, {\cal A}^a_i, \lambda, \lambda_i, \lambda_a)$
as a set of dynamical variables, then the
principal part of (\ref{lambdaC1})-(\ref{lambdaC3})
can be written as
\begin{eqnarray}
\partial_t\lambda &\cong&
 -i\alpha_1\epsilon^{bcd} \tilde{E}^j_c \tilde{E}^l_d  (\partial_l{\cal A}^b_j),
\\
\partial_t\lambda_i&\cong&
\alpha_2
[-e \delta^l_i \tilde{E}^j_b
+e \delta^j_i \tilde{E}^l_b
](\partial_l{\cal A}^b_j),
\\
\partial_t\lambda_a&\cong&\alpha_3
\partial_l\tilde{E}^l_a.
\end{eqnarray}

The characteristic matrix of the system ${u}^{(DL)}_\alpha$ does not
form a Hermitian matrix.  However,
if we modify the right-hand-side of
the evolution equation of ($\tilde{E}^i_a, {\cal A}^a_i$), 
then the set becomes
a symmetric hyperbolic system.
This is done by adding
$\bar{\alpha}{}_3 \gamma^{il}(\partial_l \lambda_a)$
to the equation of $\partial_t \tilde{E}^i_a$,
and by adding
$i\bar{\alpha}{}_1\epsilon^a{}_c{}^d \tilde{E}^c_i \tilde{E}^l_d
(\partial_l \lambda)
+
\bar{\alpha}{}_2
(-e \gamma^{lm} \tilde{E}^a_i
+e \delta^m_i \tilde{E}^{la}  )
(\partial_l \lambda_m)
$ to the equation of $\partial_t{\cal A}^a_i$.
The final principal part, then, is written as
\begin{equation}
\partial_t \left(
\matrix{\tilde{E}^i_a \cr {\cal A}^a_i \cr \lambda
 \cr \lambda_i \cr \lambda_a}
\right)
\cong
\left(
\matrix{
{\cal M}(\tilde{E},\tilde{E})^l {}_a {}^{bi}{}_m & 0 & 0 & 0&
 \bar{\alpha}{}_3 \gamma^{il}\delta_a{}^b  
\cr 
0&  {\cal M}({\cal A},{\cal A})^l{}^a{}_i{}_b{}^m&
i\bar{\alpha}{}_1\epsilon^a{}_c{}^d \tilde{E}^c_i \tilde{E}^l_d
&
\bar{\alpha}{}_2 e 
(
 \delta^m_i \tilde{E}^{la} - \gamma^{lm} \tilde{E}^a_i )
 & 0 
\cr 
0 & 
-i\alpha_1\epsilon_b{}^{cd} \tilde{E}^m_c \tilde{E}^l_d
& 0 & 0 & 0 
\cr 
0 &
\alpha_2 e 
(\delta^m_i \tilde{E}^l_b -\delta^l_i \tilde{E}^m_b)
& 0 & 0 & 0  
\cr 
\alpha_3\delta_a{}^b \delta^l{}_m& 0 & 0 & 0& 0
}
\right)
\partial_l \left(
\matrix{\tilde{E}^m_b \cr {\cal A}^b_m \cr \lambda
 \cr \lambda_m \cr \lambda_b}
\right).  
\label{DClambda-system}
\end{equation}
Clearly, the solution
$(\tilde{E}^i_a, {\cal A}^a_i, \lambda, \lambda_i, \lambda_a)
=(\tilde{E}^i_a, {\cal A}^a_i, 0, 0, 0)$ represents the original solution
of the Ashtekar system.  If the $\lambda$s decay to zero
after the evolution, then the solution also describes the original
solution of the Ashtekar system in that stage.
Since the dynamical system of ${u}^{(DL)}_\alpha$, 
(\ref{DClambda-system}),
 constitutes a symmetric
hyperbolic form, the solutions to the $\lambda$-system are unique.
BFHR showed analytically that such a decay of $\lambda$s can be seen
for $\lambda$s sufficiently close to zero
with a choice of appropriate combination of $\alpha$s and $\beta$s, 
and that statement can be
also applied to our system.
Therefore, the dynamical system, (\ref{DClambda-system}), is useful
for stabilizing numerical simulations from the point that it recovers
the constraint surface automatically.

\section{Asymptotically real-valued system}\label{sec:4}
We next extend the system, (\ref{DClambda-system}), 
to the one that also has an attractor
in the real-valued surfaces.
Since the dynamical system of ${u}^{(D)}_\alpha$ requires the
triad reality condition to form a symmetric hyperbolic system,
our
purpose is to construct a system which asymptotically evolves into
the triad-real-valued manifold.

In order to obtain such a system, we add two more new variables,
$\lambda{}^{ij}$ and $\lambda^i_a$, which satisfy the evolution equations
\begin{eqnarray}
\partial_t\lambda{}^{ij}&=&
\alpha_4 \, W^{ij}
-\beta_4 \, \lambda{}^{ij}, \label{lambdaR1}
\\
\partial_t\lambda{}^i_a&=&
\alpha_5 \, U^i_a
-\beta_5 \, \lambda{}^i_a, \label{lambdaR2}
\end{eqnarray}
corresponding to the secondary metric reality conditions,
(\ref{w-reality2-final}), and the
primary triad reality conditions, (\ref{s-reality1}), respectively.
The equation (\ref{lambdaR1}) is necessary to complete this system
since the term $W^{ij}$ appears in the non-principal part of the
equation $\partial_t U^i_a$. Thus, these two equations will guarantee
the consistency
with the secondary condition. 
The principal terms of (\ref{lambdaR1}) and (\ref{lambdaR2}) become
\begin{eqnarray}
\partial_t\lambda{}^{ij} &\cong& \alpha_4 
{\rm Re}(\epsilon^{abc}\tilde{E}^k_c
\tilde{E}^{(j}_a \partial_k \tilde{E}^{i)}_b),
\\
\partial_t\lambda{}^i_a &\cong& 0,
\end{eqnarray}
under the assumption of the triad reality condition.

A set of variables ${u}^{(DLR)}_\alpha=
(\tilde{E}^i_a, {\cal A}^a_i, \lambda, \lambda_i, \lambda_a, \lambda{}^{ij},
\lambda{}^i_a)$,
then,
forms a symmetric hyperbolic system
if we further modify the equations for $\partial_t \tilde{E}^i_a$
similarly to the case of (\ref{DClambda-system}). The final set of
equations
in the matrix form can be written as
\begin{equation}
\partial_t \left(
\matrix{\tilde{E}^i_a \cr {\cal A}^a_i \cr \lambda
 \cr \lambda_i \cr \lambda_a \cr
\lambda{}^{ij} \cr
\lambda{}^i_a}
\right)
\cong
\left(
\matrix{
&&& \bar{\alpha}_4 {\rm Re}(\epsilon_a{}^c{}_d\tilde{E}^l_c
\tilde{E}^d_{(m} \delta_{n)}^i) & 0 
\cr 
&{\cal M} (\ref{DClambda-system}) &&0  & \vdots 
\cr 
&& &\vdots &\vdots 
\cr 
\alpha_4 {\rm Re}(\epsilon^{bcd}\tilde{E}^l_c
\tilde{E}^{(i}_d \delta_m^{j)})
& 0 & \cdots & 0& 0
\cr 
0&\cdots &\cdots &0&0
}
\right)
\partial_l \left(
\matrix{\tilde{E}^m_b \cr {\cal A}^b_m \cr \lambda
 \cr \lambda_m \cr \lambda_b \cr
\lambda{}^{mn} \cr
\lambda{}^m_b}
\right), \label{DCRlambda-system}
\end{equation}
where ${\cal M} (\ref{DClambda-system})$ denotes the matrix in the
right-hand-side of (\ref{DClambda-system}).

Clearly, the solution
$(\tilde{E}^i_a, {\cal A}^a_i, \lambda, \lambda_i, \lambda_a, \lambda^{ij},
\lambda^i_a)
=(\tilde{E}^i_a, {\cal A}^a_i, \lambda, \lambda_i, \lambda_a, 0, 0)$ 
represents the original solution
of (\ref{DClambda-system}).  
The same discussion in the previous section can be applied also here. 
Therefore, we expect that the system (\ref{DCRlambda-system})
controls the violation of the triad reality condition during the time
integration. 

We remark that 
a reduced set of the variables ${u}^{(DR)}_\alpha=
(\tilde{E}^i_a, {\cal A}^a_i, \lambda{}^{ij},
\lambda{}^i_a)$ does not work for the 
purpose of controlling the reality condition, since the secondary condition
of the reality requires the constraints to be satisfied.

\section{Concluding Remarks} \label{sec:disc}

We showed a set of dynamical equations, which has 
the constraint surface as its attractor, by introducing new additional 
variables that obey dissipative equations of motion. 
Based on BFHR's analytical proof  \cite{BFHR}, we expect that
this set of equations is robust against a perturbative error of the 
constraint equations.  
Thus, the system may be useful for future numerical
studies with its stability property. 

We also showed an advanced set of equations that has its attractor also
in the real-valued surface.  Since our symmetric hyperbolic system of the
original Ashtekar's variables requires the reality condition on the triad, 
the new system is designed as such a way.  The same above discussion can be
applied to this advanced set, and we expect the asymptotically real-valued
feature in its evolution.

The problem of these systems might be that they require many 
additional variables.
{}From a view
point of numerical applications, this claim would not be so 
serious a problem, as we see a success of a dissipative 
maximal slicing condition \cite{Kdrive} (`K-driver' in the 
literature). 
Actually, the $\lambda$-system of the
Einstein equations was already tested and confirmed 
to work appropriately in numerical applications at least 
in one dimensional space evolution models \cite{private}. 
Therefore we expect our system also shows the desired
asymptotic behaviours. 
We are in preparation of presenting such a numerical result.
We are also trying to reduce the number of the variables
in order to find out clear geometrical meanings of 
our $\lambda$-system. 
We will report these efforts elsewhere.


\vspace{0.2cm}

\noindent
{\bf Acknowledgments}  ~~
We thank Abhay Ashtekar for his comments on the idea of this work.
We also thank Sean Hayward for careful reading of this manuscript.
A part of this work was done when HS was at Dept. of Physics, 
Washington University, St. Louis, Missouri. 
HS was partially supported by NSF PHYS 96-00049, 96-00507,
and NASA NCCS 5-153 when he was at WashU.
HS was supported by the Japan Society for the Promotion of Science.


\end{document}